\documentclass[preprint,12pt]{aastex}
\usepackage{natbib}
\usepackage{amssymb}

\newcommand\bmath[1] {\mbox{\boldmath$\rm #1$}}
\renewcommand{\d}{{\rm d}}
\newcommand{\e}{{\rm e}}

\newcommand{\be}{\begin{equation}}
\newcommand{\bel}[1]{\begin{equation}\label{eq:#1}}
\newcommand{\ee}{\end{equation}}
\newcommand{\bd}{\begin{displaymath}} %like \be, but doesn't put in eqn. number
\newcommand{\ed}{\end{displaymath}}   %like \ee, but doesn't put in eqn. number
\newcommand{\bea}{\begin{eqnarray}}
\newcommand{\beal}[1]{\begin{eqnarray}\label{eq:#1}}
\newcommand{\eea}{\end{eqnarray}}

\begin{document}

\title{
Oscillations of starless cores}
\author{Eric Keto\altaffilmark{1},
Avery Broderick\altaffilmark{1}, Charles J. Lada\altaffilmark{1},
Ramesh Narayan\altaffilmark{1}
}
\altaffiltext{1}{Harvard-Smithsonian Center for Astrophysics,
60 Garden Street, Cambridge MA 02138}

\begin{abstract}

If  the split, asymmetric 
molecular spectral line profiles that are seen in many starless cores are interpreted
as indicative of global collapse or expansion of the core then one possible implication
is that most starless cores have short lifetimes
on the order of the collapse or sound crossing time scale.  An alternative interpretation
of the line profiles as indicative of perturbations on an underlying equilibrium structure
leads to the opposite implication, that many cores have long lifetimes. While 
evidence suggests that some cores are collapsing on a free-fall time scale,
we show that observations of some other starless cores can be reproduced by a
model of non-radial oscillations about the equilibrium configuration of a
pressure-bounded, thermally-supported sphere (Bonnor-Ebert sphere).  We model
the oscillations as linear perturbations following a standard analysis developed
for stellar pulsations and compare the column densities and molecular spectral
line profiles predicted from a particular model to observations of the Bok
globule B68.

\end{abstract}

\section*{Introduction}

Observations of starless cores such as the Bok globule Barnard 68 (B68)
suggest that these isolated, small, dark, molecular clouds are well described as
discrete, self-gravitating, dynamical entities with an approximate internal
balance of thermal and gravitational forces \citep{Bok1948, AlvesLadaLada2001,
KetoField2005, Tafalla2004}.  In particular, although most of the cores are elliptical
or otherwise not strictly spherical, their internal density
structure is reasonably approximated by that of a hydrostatically supported sphere
confined by an external pressure.  In the isothermal case, these spheres are
generally referred to as Bonnor-Ebert (BE) spheres which are described by a
truncated solution of the Lane-Emden equation \citep{Bonnor1956}, the same
equation used to describe the internal density structure of stars
\citep{Chandrasekhar1939}.  The column density through the B68 core, derived
from observations of dust extinction \citep{AlvesLadaLada2001} azimuthally averaged,
shows a remarkable correspondence with that predicted by the Lane-Emden equation. 

Many, but not all, starless cores show non-Gaussian spectral line profiles indicating
gas velocities within the core typically at a fraction of the sound speed
\citep{Zhou1994, Wang1995, Gregersen1997, Launhardt1998,
GregersenEvans2000, LeeMyersTafalla1999, LeeMyersTafalla2001, Lada2003,
LeeMyersPlume2004, Redman2004, Redman2006}.
Many cores have spectral lines with a characteristic split, asymmetric profile that can be 
interpreted as indicating global collapse or expansion depending on the sense of the
asymmetry. The B68 core is also observed to have split asymmetric spectral line
profiles but in a pattern suggestive of both inward and outward motions in different sectors of
the cloud, reminiscent of the small-amplitude pulsations observed in stars
\citep{Lada2003, Cox1980}.  In the theory of stellar structure, such pulsations
are described as perturbations on the equilibrium solution of the Lane-Emden
equation \citep{Cox1980}.  Here we explore the possibility that the velocity
structure of the starless cores such as B68 may be due to perturbations from
oscillations about the equilibrium configuration of a Bonnor-Ebert sphere.  

We model one particular realization of non-radial oscillations chosen to match
observations of the dust extinction and molecular spectral line observations of
B68. We find that non-radial density perturbations due to oscillations can
produce elliptical shapes as are often seen in continuum observations of dust
emission and extinction from starless cores 
\citep{Ward-ThompsonMotteAndre1999, Bacmann2000, AlvesLadaLada2001}.
The rich variety of possible perturbed velocity fields due to oscillations is capable of
producing a variety of spectral line profiles and velocity patterns
that match the split asymmetric spectral line profiles characteristic of many
starless cores. Thus non-spherical shapes and complex spectral line profiles
can be found in model cores that although perturbed are still in equilibrium.

The possibility that models of clouds in perturbed equilibrium can have 
shapes and spectral
line profiles similar to those of 
non-equilibrium models of global expansion or collapse introduces some ambiguity
into the
interpretation of observations of starless cores. Some well-studied cores show
several indications of  
one-way gravitational collapse 
\citep[e.g. L1544 or L1521F,][]{Caselli2002a, Keto2004, KetoField2005,
vanderTak2005, Crapsi2004}.
Similarly from several points of view,  the core B68 appears to be stable.
However, in other cores the possibility of perturbed equilibrium raises the question of whether
inferences on the evolution and lifetimes of the starless cores 
may be drawn from limited observations. In particular, spectral line profiles that
indicate inward or outward motion may sometimes be interpreted as implying either 
a short-lived evolution ending in collapse or
dissolution or alternatively a long-lived evolution in oscillations such as a breathing mode. 

This study expands the results of our previous study of the hydrodynamics of starless
cores in which we modeled the oscillations of starless cores by numerical methods \citep{KetoField2005}. That earlier study offers some suggestions to reduce the
ambiguity of the interpretation of observations of starless cores. For example, in the gravitational
collapse of BE spheres with densities beyond critical, the inward velocities increase with
decreasing radius throughout most of the volume of the sphere. In a breathing mode
oscillation the higher velocities tend to be at larger radii toward the outside of the sphere.
Thus spectral observations that resolve a core or can measure the difference in width of
the spectral line profiles from the central and the outer regions of the core may provide 
additional information to help resolve the ambiguities.
%The earlier study was one-dimensional. Here we model the oscillations of starless cores by a standard
%perturbation analysis in three dimensions \citep[see, e.g.,][]{Cox1980}. 

\section{The Hydrodynamic Model}

The starless cores, particularly those that are of relatively low mass and 
density, such as B68, may 
be modeled as isothermal, pressure supported clouds. The temperature of the 
gas in starless cores
is set by the equilibrium between heating by absorption of cosmic rays and cooling by radio frequency
molecular line emission. In addition if the gas density is high enough ($n>10^5$ cm$^{-3}$)
the gas may be collisionally coupled to the dust  which is itself heated by
the external radiation field of starlight in the Galaxy and cooled by thermal emission
in the far infrared.
\citep{Larson1973, Larson1985, Evans2001, ShirleyEvansRawlings2002, Zucconi2001, StamatellosWhitworth2003, Goncalves2004, KetoField2005}. 
Detailed modeling of starless cores shows
that the gas temperature in all the starless cores varies by no more than 50\% about
a mean temperature of 10 K. In smaller cores such as B68, 
with masses less than a few 
$M_\odot$ and densities less than $\sim 10^5$ cm$^{-3}$, the gas is sufficiently transparent
to molecular line radiation and sufficiently poorly coupled to the dust that the
temperature variation is less than a few degrees. 
Thus for the purposes of our study of perturbations of the equilibrium structure of 
starless cores, we may approximate the gas temperature as isothermal. 
With the additional assumption 
that the equilibrium configuration of the cores is spherical, 
the unperturbed cores may be modeled
as isothermal Bonnor-Ebert spheres.

The general theory of stellar pulsations is discussed in a number of
references \citep[see, e.g.,][] {Cox1980}, and thus will only be summarized
here.  The eigenmodes of the oscillations are found by solving the
linearized mass, energy, and momentum conservation
equations subject to the constraints of regularity at the center of
mass, continuity of the gravitational potential, and the vanishing
of the comoving pressure variation at the surface.  Since
the time scale for temperature equilibrium due to dust continuum and
molecular line radiation
is short in comparison to the typical oscillation periods \citep{KetoField2005},
the perurbations, as well as the equilibrium configuration, 
may be assumed to be isothermal.  The equations for
the density and velocity perturbations are separable and the
eigenmodes may be written in terms of spherical harmonics as
%\begin{equation}
%\begin{gathered}
\begin{mathletters}
\begin{eqnarray}
&\rho_{nlm}(\bmath{r})
=
\rho_{nl}(r) \, Y_{lm}(\bmath{\hat{r}})
\label{pert_d}\\
&\bmath{v}_{nlm}(\bmath{r})
=
v^r_{nl}(r) \, \bmath{\hat{r}} Y_{lm}(\bmath{\hat{r}})
+
v^{\Omega}_{nl}(r) \, r \bmath{\nabla} Y_{lm}(\bmath{\hat{r}})\,,
\label{pert_v}
\end{eqnarray}
\end{mathletters}
%\end{gathered}
%\label{pert_qs}
%\end{equation}
where $l$ and $m$ are the standard angular quantum numbers,
$\hat{\bmath{r}}$ is the radial unit vector,
$\rho_{nl}(r)$, $v^r_{nl}(r)$ and
$v^\Omega_{nl}(r)$ are the radial eigenfunctions which define
the oscillation mode, and $n$ is a third quantum number
corresponding to the number of radial nodes in $v^r$ (see the
Appendix for more details).  The density and velocity field
is then a linear combination of the eigenmodes:
%\begin{equation}
%\begin{gathered}
\begin{mathletters}
\begin{eqnarray}
&\rho(\bmath{r})
=
\rho_0(r) + \sum_{nlm} A_{nlm} \rho_{nlm}(\bmath{r})\\
&\bmath{v}(\bmath{r})
=
\sum_{nlm} A_{nlm} \bmath{v}_{nlm}(\bmath{r})\,,
%\end{gathered}
%\end{equation}
\end{eqnarray}
\end{mathletters}
where $\rho_0(r)$ is the unperturbed equlibrium density profile and 
$A_{nlm}$ are dimensionless mode amplitudes.

\begin{figure}
\begin{center}
\includegraphics[width=5in]{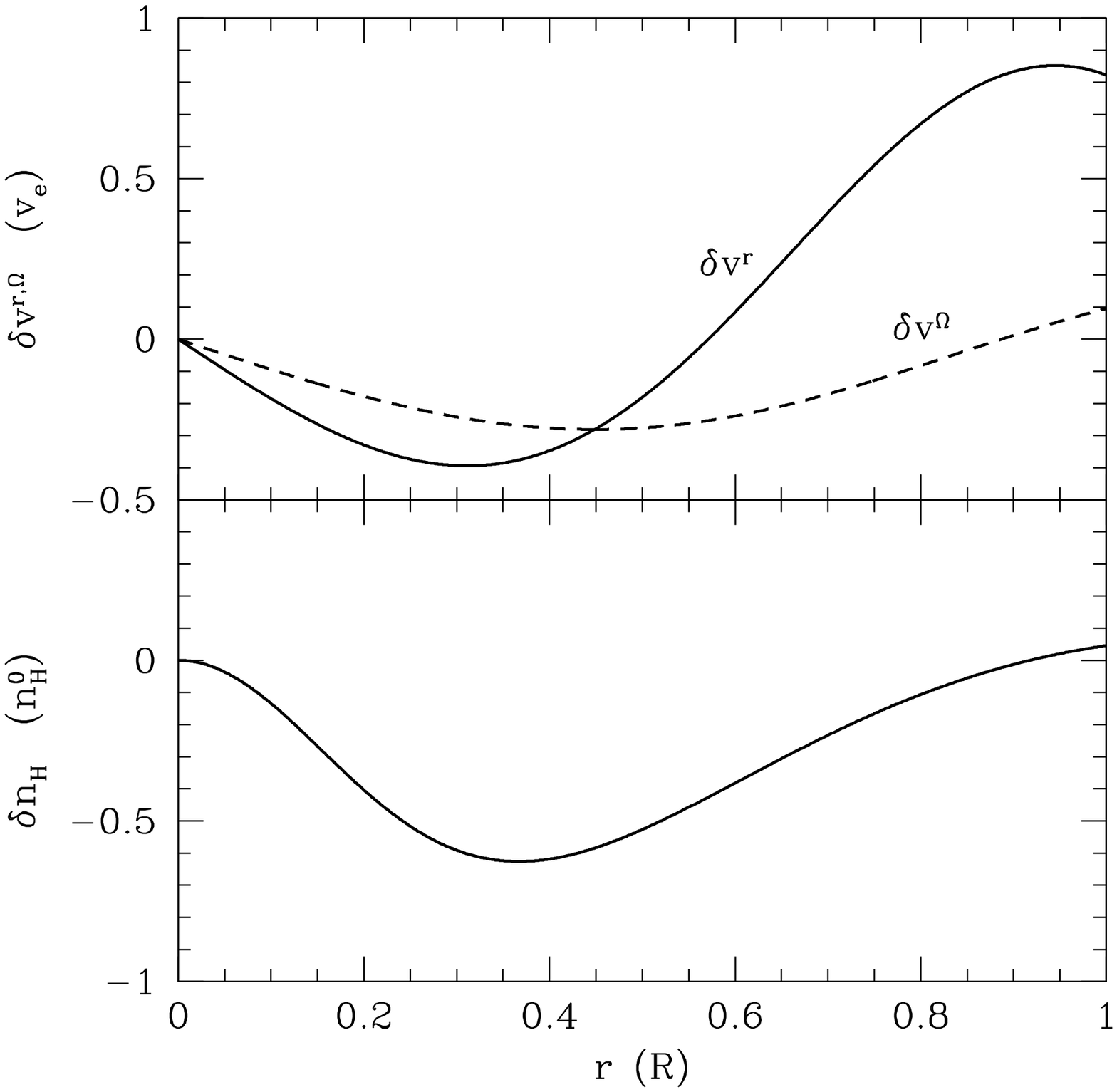}
\end{center}
\caption{The density and velocity perturbations defined by equations
  (\ref{pert_d}) and (\ref{pert_v}), respectively, for a single node, quadrapolar $p$-mode $(n=1,l=m=1)$ with a
  dimensionless amplitude of $0.25$ (as used in the model cloud).  The
  density and velocity perturbations are given in units of the central
  density ($n^0_H)\sim 4.6\times10^5\,{\rm cm^{-3}}$) and escape
  velocity ($v_e\sim 0.32\,{\rm km\,s^{-1}}$), respectively.}
\label{fig:mode12}
\end{figure}

In general a large number of modes of oscillation exist, increasing in
angular and radial complexity with increasing $n$, $l$ and $m$.  In this study, we
restrict ourselves to the simplest mode which can reproduce the
observed features of B68, i.e., one that can simultaneously produce two
velocity reversals across the cloud and a single reversal in radius.
This is the $n=1$, $l=m=2$ mode, for which the radial perturbation
eigenfunctions are shown in figure \ref{fig:mode12}.  In order to 
produce the split asymmetric line profiles characteristic of the observations,
the velocity perturbations must be a significant  fraction of the sound speed.
We chose a dimensionless amplitude
of $25\%$.  This is quite large for the assumption of linear perturbations, as can be seen in figure
\ref{fig:cdenvelope}, in which the perturbed and equilibrium density
profiles are compared. 
Nevertheless, because the calculation is intended as an example to show how a
model of oscillations could explain the observations, a linear analysis is sufficient
for the purposes.

Our model cloud is scaled to a specific mass, size, and density chosen to approximate
the properties of B68 as  derived from observations. The model cloud has
a total mass of 1.0 M$_\odot$,
a central density of $4.6\times 10^5$ cm$^{-3}$, and temperature of 10 K. 
The maximum velocity due to the oscillations is approximately 0.1 kms$^{-1}$. 

\section{Comparison with observations}

Our knowledge of the structure of starless cores comes primarily from
observations of both dust extinction and emission which trace the dust column
density, and observations of molecular spectral lines which trace the gas column
density and velocity structure of a cloud.  We calculate the predicted column
density and spectral line profiles of certain molecules for comparison with
observations.

\subsection{Column density}

\begin{figure}[t]
\begin{center}
\includegraphics[width=4.5in,angle=0]{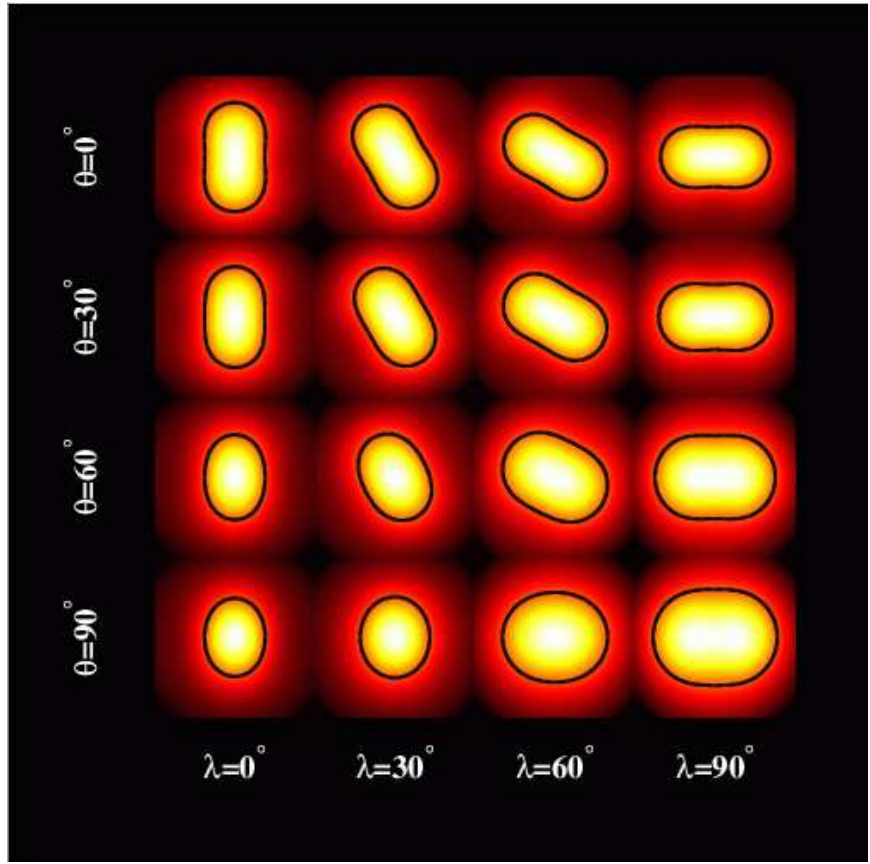}
\end{center}
\caption{Column density maps of the model cloud.  Perturbations in
  density corresponding to modes of oscillation create a non-spherical
  density distribution.  The black contour denotes the half-maximum
  column density. Latitude and longitude refer to the angles $\lambda, \theta$ of 
  a standard right-handed coordinate system with $\lambda$ in the x-y plane and
  $\theta$ off the z-axis.}
\label{fig:cdmaps} 
\end{figure}

\begin{figure}[t]
\begin{center}
\includegraphics[width=4.5in,angle=0]{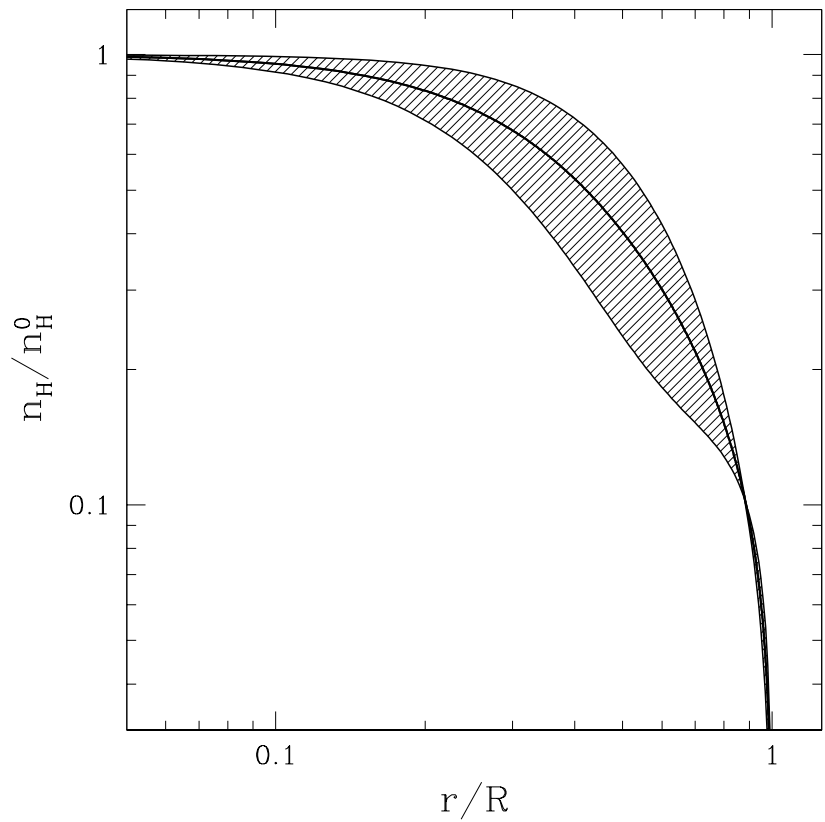}
\end{center}
\caption{Log of the column density, $n_H$, normalized by the unperturbed central density,
$n^0_H$, as a function of the log of the radius, $r$, normalized by the unperturbed radius, $R$,
as viewed along the z-axis, $\theta=0^\circ$. This orientation corresponds to the top line
of models in figures \ref{fig:cdmaps} and \ref{fig:many_views}.
The shaded portion of the figure shows the range in column density
created by the perturbations corresponding to the modes
of oscillation. The unperturbed equilibrium figure is shown as
a solid line in the middle of the range.
}
\label{fig:cdenvelope} 
\end{figure}

The column density  through the starless cores, usually derived from
observations of dust extinction or emission, is
often shown in two different presentations:  as two-dimensional
spatial surface density maps and as a radial surface density profiles usually
azimuthally averaged around the peak of the column density distribution.
The two-dimensional map of gas column
density of our perturbed model sphere is shown for various viewing orientations
in figure \ref{fig:cdmaps}.  The viewing angles $\lambda,\theta$ are
 those of a standard right-handed
coordinate system with $\lambda$ in the x-y plane and $\theta$ off the z-axis. 
The observer looks from the positive x-axis.  As can be
seen in figure \ref{fig:cdmaps}, a cloud with perturbations of sufficient
amplitude may show a morphology with an aspect ratio in column density of about
2:1 depending on viewing angle.  The column density as a function of distance
from the cloud center is shown in figure \ref{fig:cdenvelope}.  The shaded
envelope shows the range of variation of the run of column density with distance
for different directions about the cloud center.  The solid line in the middle
of the shaded envelope shows the equilibrium configuration.  Figures
\ref{fig:cdmaps} and \ref{fig:cdenvelope} may be compared to maps and plots of
the column density as in the observations of \cite{AlvesLadaLada2001},
\cite{Bacmann2000}, and \cite{Ward-ThompsonMotteAndre1999}.

\subsection{Spectral lines}

Molecular spectral line profiles as would be observed from the model
core have been computed by  
a non-LTE accelerated $\Lambda$
iteration (ALI) radiative transfer code
\citep{Keto2004}. 
The core is assumed to be at 150 pc,
the distance to the Taurus star-forming region and assumed to be
observed with a telescope with a beam width of 20$^{\prime\prime}$.
The radiative transfer calculations are done in three dimensions to capture
the asymmetric structure of the density and velocity perturbations.

To compute the spectral line profiles we also need to know
the gas phase abundances of our tracer molecules.
In the cold dense interiors of the dark cloud cores,
molecules freeze out of the gas phase onto the surfaces of dust grains 
at different characteristic densities according to their different volatilities
\citep{BrownCharnleyMillar1988, WillacyWilliams1993, HasegawaHerbstLeung1992, HasegawaHerbst1993, Caselli1999, Caselli2002a, Caselli2002b, Bergin2002, Tafalla2002, BerginLangerGoldsmith1995, Aikawa2001, Caselli2002c}.  
We follow \cite{Bergin2001}, \cite{Tafalla2002}, and \cite{Tafalla2004} and approximate 
the dependence of the abundance
on the density, $n$, as an exponential. 
\be
X(n) = X_0 \exp (-n/n_d)
\ee
For CS we use an undepleted abundance of $X_0 = 6 \times 10^{-8}$ and critical
density of $n_d = 10^{4}$ cm$^{-3}$.  
%For the extremely volatile molecule,
%N$_2$H$^+$, we assume a constant 
%abundance of $6\times 10^{-9}$.
A more complete treatment of the molecular abundances of the starless cores derived from detailed
chemical modeling is discussed in Lee, Bergin \& Evans (2004). 

\begin{figure}[t]
\begin{center}
\includegraphics[width=7.5in,angle=90]{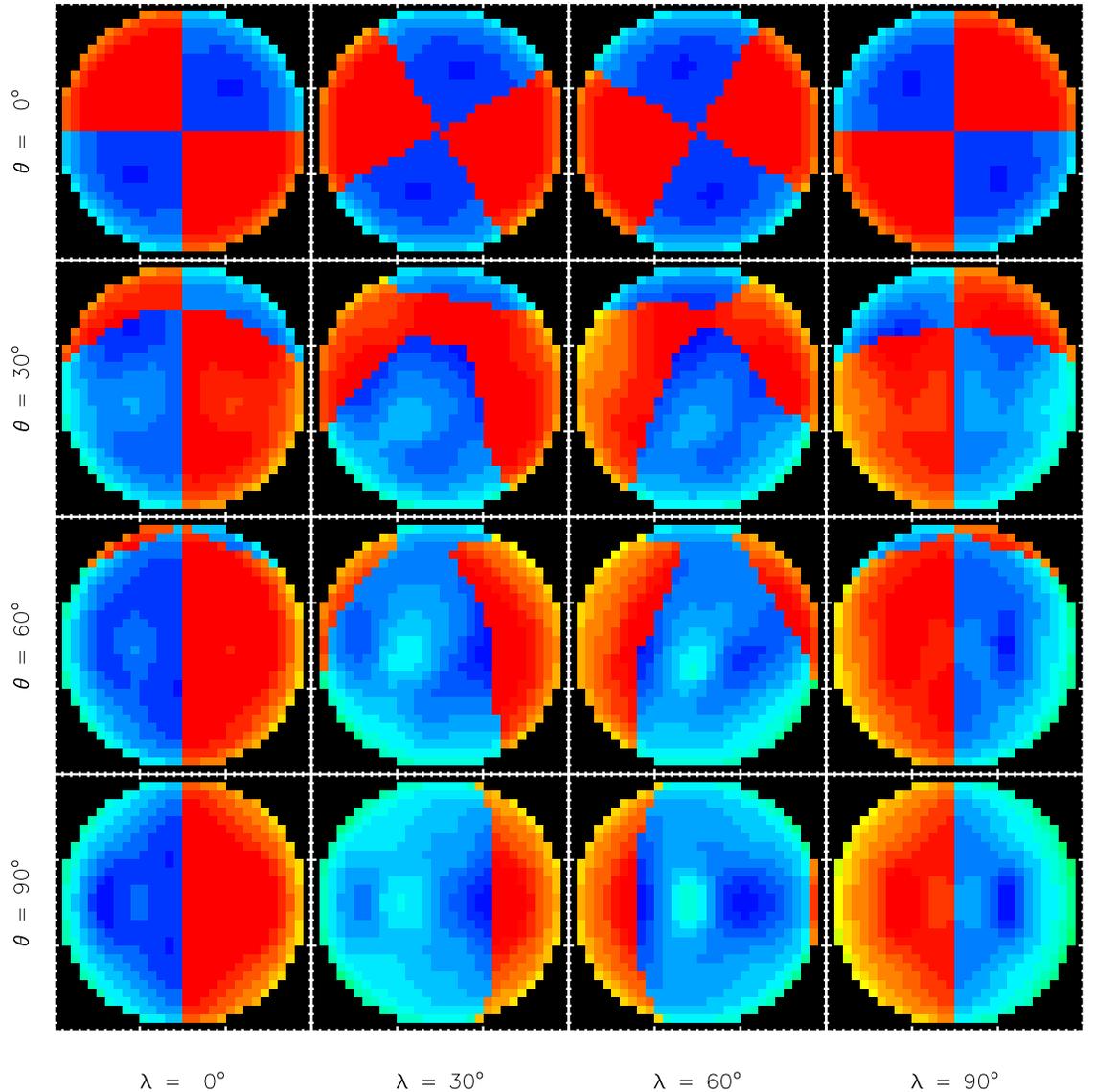}
\end{center}
\vskip -1.0truein
\caption{Map of the velocity of the peak of the CS(1-0) emission for the
model cloud at several viewing angles  $(\lambda,\theta)$. 
The viewing angles are the same as figure \ref{fig:cdmaps}.
The color scale shows peak velocities 
in the range of -0.075 kms$^{-1}$ (red) to 0.075 kms$^{-1}$ (blue). 
In these asymmetric split spectra, the brightest emission is generated in the 
rear hemisphere of the core. Thus the definition
of red as indicative of a peak emission with negative velocity  
corresponds to the convention of red as indicating contraction in the forward 
hemisphere of the core, the same convention as in
\cite{Lada2003}.
}
\label{fig:many_views} 
\end{figure}

\begin{figure}[t]
{\hskip -1.0truein\includegraphics[width=7.0in,angle=90]{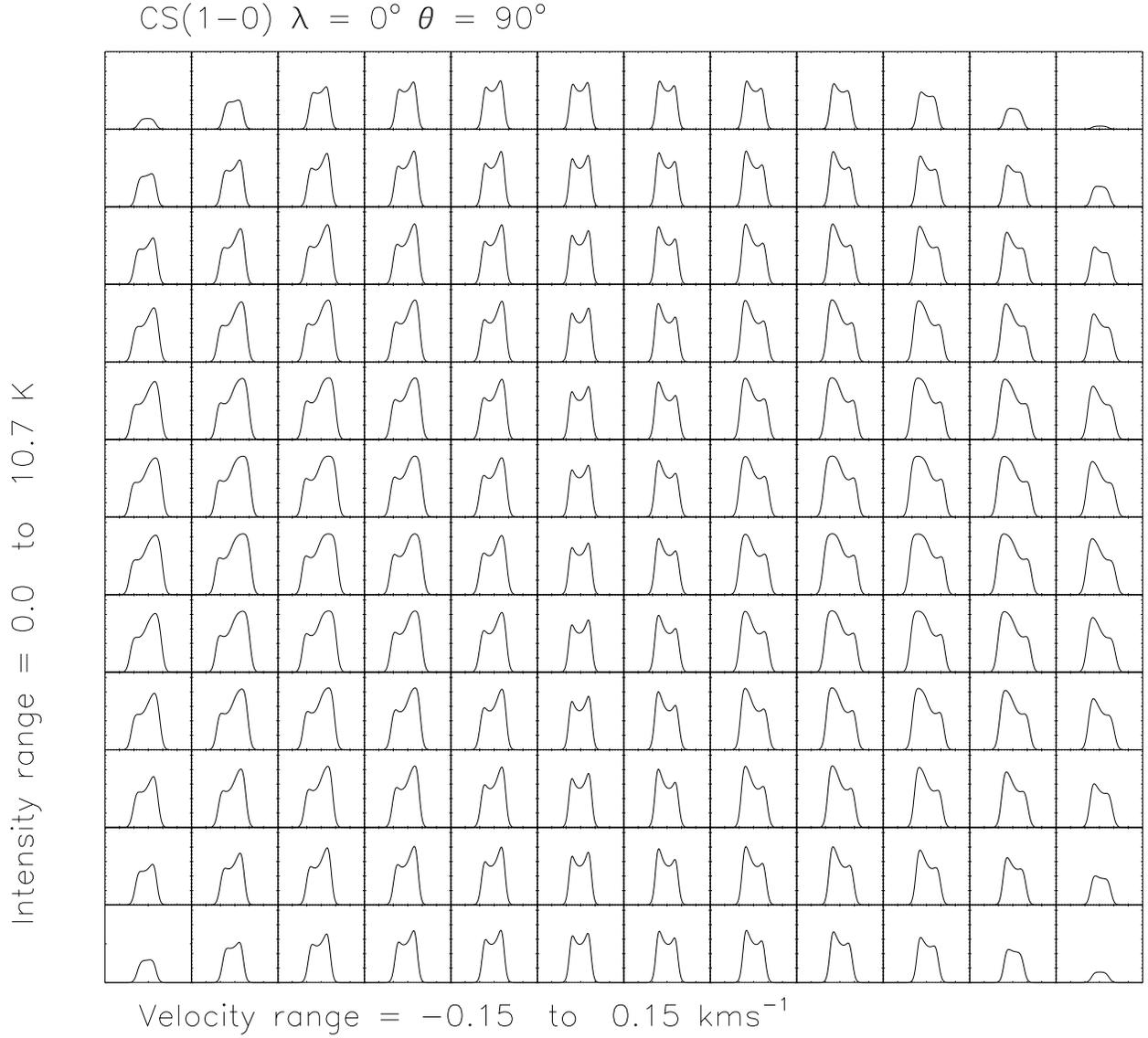}\hfill}
\caption{Spectra across the model cloud at a viewing angle of $\lambda,\theta = 0^\circ, 90^\circ$. A 
$12\times 12$
grid of spectra of CS(1-0) covering the inner 80\% of the cloud.
The velocity range is $\pm0.2$ kms$^{-1}$ and the intensity range
is 8 K.  The reversal in sign of the motions in this phase of oscillation creates spectral
line profiles that are similar to those produced by
rotation even though this cloud has no angular momentum.
}
\label{fig:mv0} 
\end{figure}

\begin{figure}[t]
{\hskip -1.0truein\includegraphics[width=7.0in,angle=90]{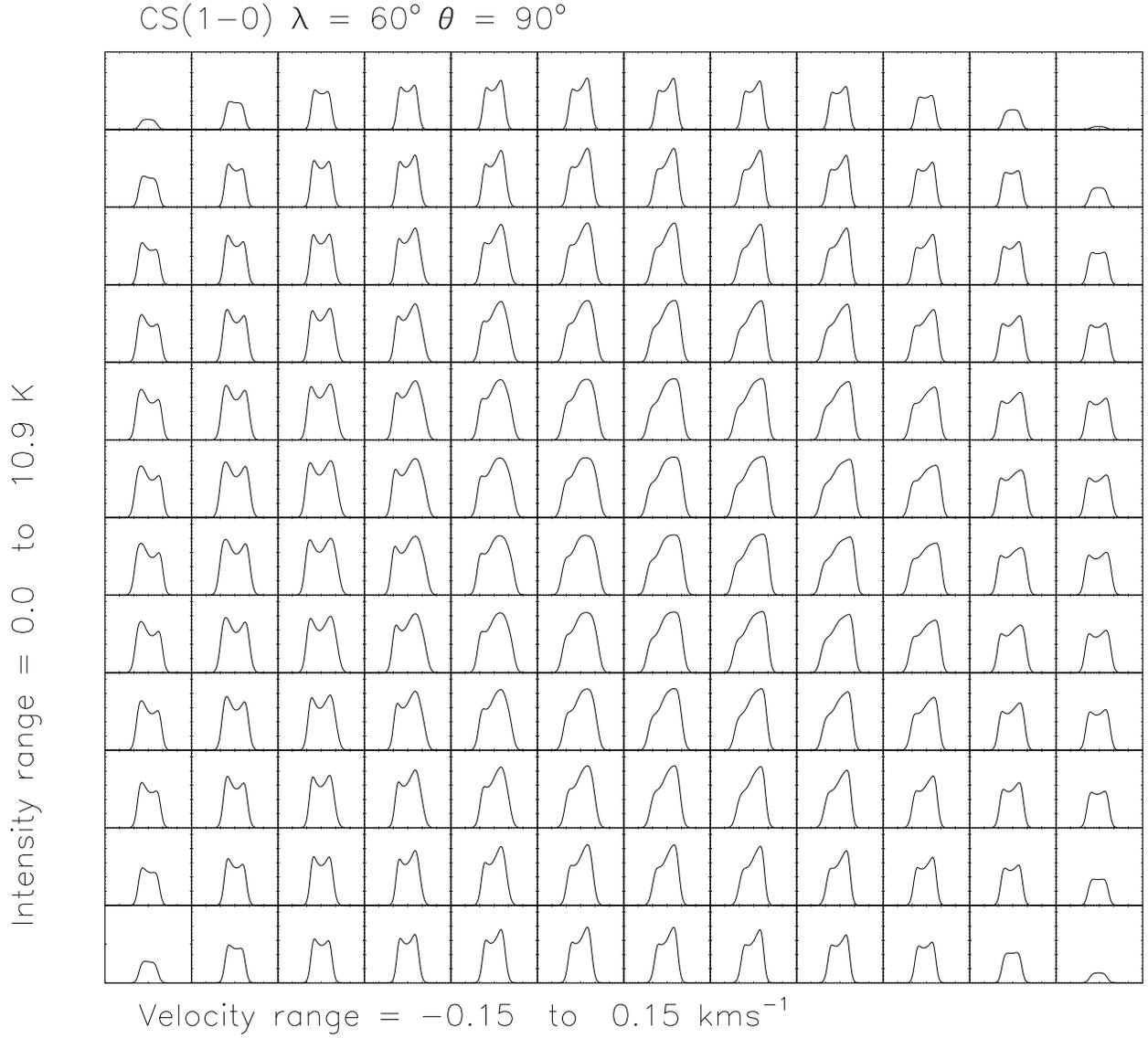}\hfill}
\caption{Spectra across the model cloud at a viewing angle of $\lambda,\theta = 60^\circ, 90^\circ$
in the same format as in figure \ref{fig:mv0}.
At this viewing angle, the pattern of spectral line profiles is similar to
that produced by continuous expansion.
}
\label{fig:mv4} 
\end{figure}

\begin{figure}[t]
{\hskip -1.0truein\includegraphics[width=7.0in,angle=90]{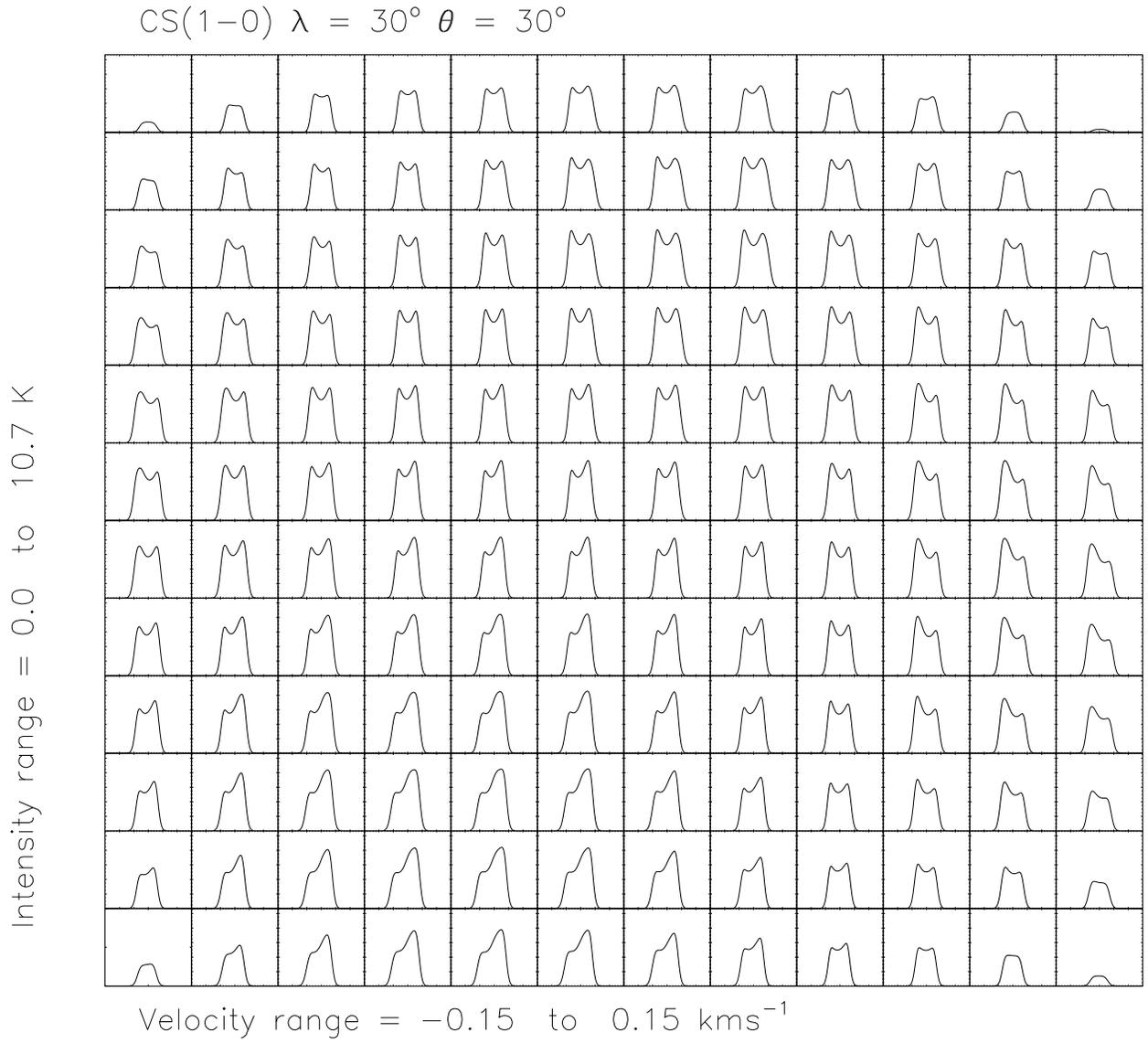}\hfill}
\caption{Spectra across the model cloud at a viewing angle of $\lambda,\theta = 30^\circ, 30^\circ$
in the same format as in figure \ref{fig:mv0}.
At this viewing angle, the pattern of spectral line profiles is similar to
that seen in the Bok globule B68.
}
\label{fig:mv30} 
\end{figure}

Figure \ref{fig:many_views} depicts the velocity of the peak of the profile of the
CS(1-0) spectral line
for the same viewing orientations as in figure \ref{fig:cdmaps}. This presentation of
the spectral line velocity is the same as in the  observations of B68
\cite[figures 6 and 7 of ][]{Lada2003}. In particular, the map at the orientation
$\lambda=30^\circ,\theta=30^\circ$ is  similar to 
that observed in B68 in that the modeled and observed 
velocity maps show two reversals of sign across
the face of the cloud. 
Such double reversals cannot be explained by models
of simple rotation, contraction, or expansion.
The velocity fields of some of the orientations shown in
figure \ref{fig:many_views} are similar to those
produced by models of simple infall, contraction, and rotation. For example, ignoring the
very weak emission at the boundary of the cloud,
the map at $\lambda=0^\circ,\theta=90^\circ$ has a velocity pattern similar to that produced
by rotation, while that at $\lambda=60^\circ,\theta=90^\circ$  
appears similar to the velocities of a continuous expansion leading to dissolution. Because the pattern
of perturbations in the mode (1,2,2) model has a two-fold symmetry about the z-axis ($\theta=0$),
 a view of the model at $\lambda=150^\circ,\theta=90^\circ$ would  appear the same as that
at $\lambda=60^\circ,\theta=90^\circ$ (expansion) except that the velocities would be
reversed and would therefore appear similar to the velocities of continuous contraction as in the
early stages of gravitational collapse.
Figures \ref{fig:mv0} and  \ref{fig:mv4}
show the spectral line profiles 
for the viewing orientations with $\lambda=0^\circ,\theta=90^\circ$ and
$\lambda=60^\circ,\lambda=90^\circ$ which are similar to those of 
rotation and expansion, or
contraction if the velocities of the latter figure are reversed.
These maps of spectral line profiles may be compared with the maps of
observed spectral line profiles from starless cores which have been
interpreted as dominated by rotation \citep[figure 4 of ][]{Redman2004} 
and infall \citep[figure 2 of][]{Williams1999}. Figure \ref{fig:mv30} shows
the spectral line profiles for the orientation with $\lambda=30^\circ,\theta=30^\circ$ which
is similar to the pattern observed in B68.

When viewed from different orientations, our model for B68 shows different patterns for the
velocities of the emission peak and different aspect ratios in column density.  The
prevalence of each of these patterns in an ensemble of identical clouds is determined by the
symmetry of the particular mode of oscillation.  Because our chosen model for B68 with mode (1,2,2)
oscillations has a two-fold symmetry, at viewing angles along the equator the cloud will
appear to be primarily contracting or expanding on views separated by $90^\circ$.  
At intermediate viewing angles, the cloud will show a single reversal in
the observed velocity and in the sense of asymmetry in the spectral line profile.  Viewed at
intermediate latitudes between the equator and the pole, the cloud will show more complex
velocity patterns, for example the two reversals seen in figure \ref{fig:mv30}. 
These simple considerations suggest that for the (1,2,2) mode, velocity patterns that are
similar to those of radial flow, either contraction or expansion, should occur at roughly
40\% of the viewing angles. Patterns similar to rotation would be seen in another 40\% of the
angles with complex patterns as seen in B68 making up the remaining 20\%.  While these
quantitative estimates apply only to the particular (1,2,2) mode, the implication is that 
complex motions such as those seen in B68 might be seen in a minority, but still significant, 
number of observations of randomly oriented cores.

Oscillations of molecular clouds have a very long damping time and therefore should be
long-lived and easily observable.  Hydrodynamic models of one-dimensional oscillations in
starless cores indicate that the damping time due to radiative losses is many crossing times
\citep{KetoField2005}.  Thus unless the clouds have lifetimes of many tens of crossing times
and remain unperturbed for most of their lives, observations should commonly detect
oscillations of starless cores. Oscillations that are of low order should be more commonly
identified in observations than high order oscillations. First, cores dominated by  low order
oscillations would be more prevalent because the low order modes dissipate more slowly. 
Second, the spectral line shapes and patterns of the low order modes would be more easily
recognized as coherent patterns and less likely
to be confused with observational noise.

\section{Conclusions}

Our study suggests that oscillating,
otherwise stable, starless cores may be more common than previously supposed.
The density and velocity structure produced by a model of oscillations within a molecular
cloud in pressure-supported equilibrium  explains many of the observations of starless
cores.  In particular,  an oscillation model can produce self-reversed, asymmetric line
profiles with differing patterns of contraction and expansion across the face of a cloud core.  These
patterns can include  simple spatial variations such as spherically symmetric contraction
or expansion as well as more complex patterns with spatially alternating
infall and outflow signatures.  The nature and complexity of the pattern depends on both the
orientation of the core to the line-of-sight and the complexity of the mode (or modes) of the
oscillation.  Thus, globules or starless cores in oscillation can be characterized by velocity fields
that otherwise mimic global infall (contraction) or expansion  or even simple rotation.  The
implication is that some cores that have been thought to be collapsing or rotating
may be instead oscillating cores in disguise.  Indeed, our modelling indicates that if
oscillating cores in equilibrium are relatively common, then some starless cores should be
found to display global expansion that would be otherwise difficult to understand.
Moreover, additional examples of cores with alternating contraction and expansion patterns
across the face of the core, such as
B68, should also be detected in future surveys.    Such starless
cores may have lifetimes significantly longer than their sound crossing times.

\appendix
\section{Non-Radial Isothermal Oscillation Modes}
General non-radial oscillations are discussed in a number of
references (see, e.g., Cox, 1980), and thus the method by which we
constructed the isothermal oscillation modes are only summarized
here.  The equations governing small, non-radial oscillations are the
linearized mass, momentum and energy conservation equations.  Prior to
presenting these, it is necessary to be explicit about the type of
perturbation under consideration.  Eulerian perturbations at a position
$\mathbf{r}$ are defined relative to the stationary unperturbed quantity,
\begin{equation}
f' = f(\bmath{r},t)-f_0(\bmath{r},t)
\quad\Rightarrow\quad
\frac{\partial f'}{\partial t} = \left(\frac{\partial f}{\partial t}\right)',\,
\bmath{\nabla} f' = \left( \bmath{\nabla} f \right)'\,,
\end{equation}
and commute with time and spatial derivatives.  Alternatively,
Lagrangian perturbations are defined relative to the advected
unperturbed quantity,
\begin{equation}
\delta f = f(\bmath{r},t) - f_0(\bmath{r}_0,t)
\quad\Rightarrow\quad
\frac{\d \delta f}{\d t} = \delta \frac{\d f}{\d t}\,,
\end{equation}
(where $\bmath{r}_0$ is the unperturbed positions of the fluid element) which commute with the
convective derivative $\d/\d t$.  These are not independent, related by
\begin{equation}
f' = \delta f - \delta\bmath{r}\cdot\bmath{\nabla} f_0\,,
\end{equation}
and hence linear Lagrangian perturbations correspond to linear
Eulerian perturbations and vice-versus.

Energy conservation gives generally
\begin{equation}
\frac{\d \ln P}{\d t}
=
\Gamma_1 \frac{\d \ln \rho}{\d t}
+
\frac{\rho}{P}\left(\Gamma_3-1\right) \frac{\d q}{\d t}\,,
\end{equation}
where $dq/dt$ is the heating rate and 
\begin{equation}
\Gamma_1 \equiv \left(\frac{d \ln P}{d \ln \rho}\right)_s
\qquad{\rm and}\qquad
\Gamma_3 \equiv 1 + \left(\frac{d \ln T}{d \ln \rho}\right)_s\,,
\end{equation}
are given by their standard definitions (in which $s$ is the specific
entropy).  The perturbed energy equation is given to linear order by
\begin{equation}
\frac{1}{P_0} \frac{\d\delta P}{\d t}
=
\frac{\Gamma_1}{\rho_0} \frac{\d\delta\rho}{\d t}
+
\frac{\rho_0}{P_0}\left(\Gamma_3-1\right)
\frac{\d\delta q}{\d t}\,,
\end{equation}
where it was assumed that the equilibrium configuration is stationary
(eq. 5.34a in \citealt{Cox1980}).
>From the ideal gas law it is straightforward to show that the
Lagrangian perturbation to the heat flow required to maintain a
constant temperature in spite of compressional heating is given by 
\begin{equation}
\delta q
=
- \frac{(\Gamma_1-1)}{(\Gamma_3-1)} \frac{P_0}{\rho_0^2} \delta\rho\,,
\end{equation}
and hence the linearized energy equation may be integrated to yeild
\begin{equation}
\frac{\delta P}{P_0} = \frac{\delta \rho}{\rho_0}
\quad\Rightarrow\quad
\frac{P'}{P_0} = \frac{\rho'}{\rho_0}\,,
\end{equation}
for oscillations in an isothermal ideal gas.

The linearized continuity and Euler equations are given by
\begin{equation}
\delta\rho + \rho_0 \bmath{\nabla}\cdot\delta\bmath{r} = 0
\qquad{\rm and}\qquad
\frac{\d^2 \delta\bmath{r}}{\d t^2} = -
\delta\left(\frac{\bmath{\nabla} P}{\rho}\right)
+ \delta \left( \bmath{\nabla} \psi\right)\,,
\end{equation}
where $\psi$ is the gravitational potential
(eqs. 5.29a \& 5.31 in \citealt{Cox1980}).  These may be combined with
the energy equation and the first law of thermodynamics to yield
%\begin{equation}
%\begin{aligned}
\begin{eqnarray}
\frac{\d^2 \delta\bmath{r}}{\d t^2}
=&&
-
\bmath{\nabla} \left( \frac{P'}{\rho_0} + \psi' \right)
+
\left(\Gamma_1 \bmath{\nabla} \ln \rho_0 - \bmath{\nabla} \ln P_0 \right)
\frac{P_0}{\rho_0} \bmath{\nabla}\cdot\delta\bmath{r}
-
\left(\Gamma_3-1\right) \delta q \bmath{\nabla} \ln\rho_0\nonumber\\
=&&
-
\bmath{\nabla} \left( \frac{P'}{\rho_0} + \psi' \right)
-
\left(\bmath{\nabla} \frac{P_0}{\rho_0} \right)
\bmath{\nabla}\cdot\delta\bmath{r}\,.
%\end{aligned}
\label{eq:mode_eq}
%\end{equation}
\end{eqnarray}
This is simply the equation for the non-radial adiabatic oscillations
of a $\Gamma_1 = 1$ gas \citep[cf.][]{Cox1980}, which is not
unexpected as adiabatic varations of the $\Gamma_1=1$ gas are also
isothermal.  Coupled with the Poisson equation for the gravitational
potential, this provides four equations for the components of
$\delta\bmath{r}$, and $\psi'$.

In practice, it is more convenient to solve for the Dziembowski
variables \citep{Dziembowski1971}:
\begin{equation}
\bmath{\eta} =
\left(\frac{\delta r}{r} ,
\frac{P'+\rho_0 \psi'}{\rho_0 g r} ,
\frac{\psi'}{g r} ,
\frac{1}{g}\frac{\partial \psi'}{\partial r}
\right)\,,
\end{equation}
where $g \equiv -\partial\psi_0/\partial r$ is the gravitational
acceleration.  Since equation (\ref{eq:mode_eq}) is separable in
angle, it is useful to expand the Dziembowski variables in spherical
harmonics, i.e.,
\mbox{$\bmath{\eta}(\bmath{r})=\bmath{\eta}(r)\e^{i\omega t}Y_{lm}(\hat{\bmath{r}})$}.
In terms of the $\bmath{\eta}(r)$, equation (\ref{eq:mode_eq}) and the
Poisson equation are given by
%\begin{subequations}
%\begin{equation}
\begin{mathletters}
\begin{eqnarray}
&&r \frac{\partial \eta_1}{\partial r}
=
(V-3) \eta_1 + \left[\frac{l(l+1)}{\sigma^2 C} - V\right] \eta_2 + V \eta_3
\label{eq:dz_a}\\
%\end{equation}
%\begin{equation}
&&r \frac{\partial \eta_2}{\partial r}
=
\sigma^2 C \eta_1 + (1-U)\eta_2\\
%\end{equation}
%\begin{equation}
&&r \frac{\partial \eta_3}{\partial r}
=
(1-U) \eta_3 + \eta_4\\
%\end{equation}
%\begin{equation}
&&r \frac{\partial \eta_4}{\partial r}
=
U V \eta_2 + \left[ l(l+1) - UV \right] \eta_3 - U \eta_4\,,
%\end{equation}
\label{eq:dz_eqs}
%\end{subequations}
\end{eqnarray}
\end{mathletters}
where the dimensionless Chandrasekhar structure variables
($U$, $V$, $C$ and $\sigma^2$) are defined in terms of the underlying
equilibrium state by
\begin{equation}
U \equiv \frac{\partial \ln r^2 g}{\partial \ln r}
\,,\quad
V \equiv - \frac{\partial \ln P_0}{\partial \ln r}
\,,\quad
C \equiv -\frac{G M}{r^2 g} \left(\frac{r}{R}\right)^3
\,,\quad
\sigma^2 \equiv \frac{R^3\omega^2}{G M}\,,
\end{equation}
in which $M$ and $R$ are the mass and radius of the core, respectively.

These must be supplemented with boundary conditions.  Regularity at
the origin gives
\begin{equation}
\sigma^2 C \eta_1 \big|_{r=0} = l \eta_2 \big|_{r=0}
\qquad{\rm and}\qquad
\eta_4\big|_{r=0} = l \eta_3\big|_{r=0}\,.
\end{equation}
Continuity of the gravitational acceleration at the surface gives
\begin{equation}
\eta_4\big|_{r=R} = -(l+1) \eta_3\big|_{r=R}\,.
\end{equation}
Finally, the requirement that the Lagrangian variation in the pressure
at the surface vanish gives
\begin{equation}
\eta_2\big|_{r=R} = \eta_1 + \eta_3\big|_{r=R}\,.
\end{equation}
These boundary conditions together with the equations
(\ref{eq:dz_eqs}) define an eigenvalue problem for the
$\bmath{\eta}(r)$.

Given the solutions to equations (\ref{eq:dz_a}--\ref{eq:dz_eqs}), it is possible to
reconstruct the physically interesting perturbed quantities.  In terms
of the Dziembowski variables, the perturbation velocity is given by
\begin{equation}
\delta\bmath{v}(\bmath{r}) = \frac{\d \delta\bmath{r}}{\d t}(\bmath{r})
=
-i \omega \eta_1(r) \bmath{r} Y_{lm}(\hat{\bmath{r}})
-
i \omega \frac{r^2}{\sigma^2 C} \eta_2(r) \bmath{\nabla} Y_{lm}(\hat{\bmath{r}})\,,
\end{equation}
where the factor of $i$ is due to the $\pi/2$ phase shift between the
velocity and displacement.  The Eulerian density perturbation may be
obtained from this using the linearized continuity equation, which
after some simplification is
\begin{equation}
\rho'(\bmath{r}) = \rho_0 V \left[ \eta_2(r) - \eta_3(r) \right] Y_{lm}(\hat{\bmath{r}})\,.
\end{equation}

\clearpage

\begin{thebibliography}

\bibitem[Alves, Lada, \& Lada(2001)]{AlvesLadaLada2001}
   Alves, J., Lada, C., \& Lada, E. 2001, \nat, 409, 159

\bibitem[Aikawa et al.~(2001)]{Aikawa2001}
   Aikawa, Y., Ohashi, N., Inutsuka, S., Herbst, E., \& Takakuwa, S.
   2001 Ap J, 552, 632
   
\bibitem[Bacmann et al.~(2000)]{Bacmann2000}
   Bacmann, A., Andre, A.P., Puget, J.-L., Abergel, A., Bontemps, S., \&
   Ward-Thompson, D. 2000, A\&A, 361, 555
   
 \bibitem[Bergin, Langer, \& Goldsmith(1995)]{BerginLangerGoldsmith1995}
   Bergin, E., Langer, W., \& Goldsmith, P. 1995, Ap J, 441, 222
   
\bibitem[Bergin et al.~(2001)]{Bergin2001}
  Bergin, E., Ciardi, D., Lada, C., Alves, J., Lada, E., 2001, Ap J, 557, 209

\bibitem[Bergin et al.~(2002)]{Bergin2002}
   Bergin, E., Alves, J., Huard, T., \& Lada, C. 2002, Ap Jl, 570, L101
                                                                                
\bibitem[Bok(1948)]{Bok1948}
  Bok, B., 1948, {\it Centennial Symposia}, Harvard Observatory Monographs No. 7, 
  Harvard-College Observatory, Cambridge                                                                                
   
\bibitem[Bonnor(1956)]{Bonnor1956}
   Bonnor, W., 1956, MNRAS, 116, 351

\bibitem[Brown, Charnley, \& Millar(1988)]{BrownCharnleyMillar1988}
    Brown, P., Charnley, S., \& Millar, T., 1988, MNRAS, 231, 409
    
\bibitem[Caselli et al.(1999)]{Caselli1999}
   Caselli, P., Walmsley, C., Tafalla, M., Dore, L., \&
   Myers, P. 1999, Ap Jl, 523, 165

\bibitem[Caselli et al.(2002a)]{Caselli2002a}
   Caselli, P., Walmsley, C., Zucconi, A., Tafalla, M., Dore, L.,
   \& Myers, P. 2002a, Ap J, 565, 331

\bibitem[Caselli et al.(2002b)]{Caselli2002b}
   Caselli, P., Walmsley, C., Zucconi, A., Tafalla, M., Dore, L.,
   \& Myers, P. 2002b, Ap J, 565, 344

\bibitem[Caselli et al.(2002c)]{Caselli2002c}
   Caselli, P., Benson, P., Myers, P., \& Tafalla, M.,
   2002c, Ap J, 572, 238

\bibitem[Chandrasekhar(1939)]{Chandrasekhar1939} 
  Chandrasekhar, S., 1939, Stellar Structure, Univ. Chicago Press 

\bibitem[Cox(1980)]{Cox1980}
  Cox, J.~P., 1980, Theory of Stellar Pulsation, Princeton University
  Press, Princeton, NJ, Princeton
  
  \bibitem[Crapsi et al.(2004)]{Crapsi2004}
  Crapsi, A., Caselli, P., Walmsley, C., Tafalla, M., Lee, C., Bourke, T., Myers, P., 2004,
  AA 420, 957

\bibitem[Dziembowski(1971)]{Dziembowski1971}
  Dziembowski, W., 1971 Acta Astron., 21, 289

\bibitem[Evans et al.~(2001)]{Evans2001}
   Evans, N., Rawlings J.,  Shirley, Y., \& Mundy, L.
   2001, Ap J, 557, 193                                                                                
  
\bibitem[Hasegawa, Herbst, \& Leung(1992)]{HasegawaHerbstLeung1992}
    Hasegawa, T., Herbst, E. \& Leung, C., 1992, Ap Js, 82, 167

\bibitem[Hasegawa \& Herbst(1993)]{HasegawaHerbst1993}
    Hasegawa, T., \& Herbst, E., 1993, MNRAS, 261, 83


\bibitem[Goncalves, Galli, \& Walmsley(2004)]{Goncalves2004}
   Goncalves, J., Galli, D., Walmsley, M., 2004, A\&A, 415, 617

\bibitem[Gregersen et al.~(1997)]{Gregersen1997}
  Gregersen, E., Evans, N., Zhou, S., Choi, M., 1997, Ap J, 484, 256

\bibitem[Gregersen \& Evans(2000)]{GregersenEvans2000}
  Gregersen, E., Evans, N., 2000, Ap J, 538, 260

\bibitem[Keto \& Field(2005)]{KetoField2005}\
	Keto, E. \& Field, G., 2005, ApJ in press, astro-ph/0508527
	
\bibitem[Lada et al.(2003)]{Lada2003}
	Lada, C., Bergin, E., Alves, J., Huard, T., 2005, Ap J 586, 286

\bibitem[Launhardt et al.~(1998)]{Launhardt1998}
  Launhardt, R., Evans, N., Wang, Y., Clemens, D., Henning, T., Yun, J., 1998, Ap Js, 119, 59

\bibitem[Larson(1973)]{Larson1973}
   Larson, R. 1973, Fundam. Cosmic Phys., 1, 1

\bibitem[Larson(1985)]{Larson1985}
   Larson, R. 1985, MNRAS, 214, 379

\bibitem[Lee, Bergin \& Evans(2004)]{LeeBerginEvans2004}
	Lee, J., Bergin, E., Evans, N., 2004, Ap J, 617, 360

\bibitem[Lee \& Myers(1999)]{LeeMyers1999}
   Lee, C.W. \& Myers, P.C. 1999, Ap Js, 123, 233

\bibitem[Lee, Myers, \& Plume(2004)]{LeeMyersPlume2004}
  Lee, C.W., Myers, P., Plume, R., 2004, Ap Js, 153, 253

\bibitem[Lee, Myers, \& Tafalla(1999)]{LeeMyersTafalla1999}
   Lee, C., Myers, P., \& Tafalla, M. 1999, Ap J,  576, 788

\bibitem[Lee, Myers, \& Tafalla(2001)]{LeeMyersTafalla2001}
   Lee, C., Myers, P., \& Tafalla, M. 2001, Ap Js,  136, 703

\bibitem[Keto et al.(2004)]{Keto2004}
    Keto, E., Rybicki, G., Bergin, E., Plume, R., 2004, Ap J, 613, 355


\bibitem[Redman et al.(2004)]{Redman2004}
	Redman, M., Keto, E., Rawlings, J., Williams, D., 2004, MNRAS, 352, 1365
	
\bibitem[Redman et al.(2006)]{Redman2006}
Redman, M., Keto, E., Rawlings, J., Williams, D., 2006, MNRAS, submitted

\bibitem[Shirley, Evans, \& Rawlings(2002)]{ShirleyEvansRawlings2002}
   Shirley, Y., Evans, N.,  \& Rawlings, J., 2002 Ap J, 575, 337

\bibitem[Stamatellos \& Whitworth(2003)]{StamatellosWhitworth2003}
   Stamatellos, D., \& Whitworth, A., 2003, A\&A, 407, 941

\bibitem[Tafalla et al.~(1998)]{Tafalla1998}
  Tafalla, M., Mardones, D., Myers, P., Caselli, P., Bachiller, R., Benson, P.,
  1998, Ap J, 504, 900
   
\bibitem[Tafalla et al.~(2002)]{Tafalla2002}
   Tafalla, M., Myers, P., Caselli, P., Walmsley, C., \& Comito, C.
            2002, Ap J, 569, 815   

\bibitem[Tafalla et al.~(2004)]{Tafalla2004}
   Tafalla, M., Myers, P., Caselli, P., Walmsley, M., 2004, A\&A, 416, 191

\bibitem[van der Tak et al.(2005)]{vanderTak2005}
van der Tak, F., Caselli, P., Ceccarelli, C., 2005, AA, 439, 135                                                                      


\bibitem[Wang et al.~(1995)]{Wang1995}
  Wang, Y., Evans, N., Shudong, Z., Clemens, D., 1995, Ap J, 454, 742

\bibitem[Ward-Thompson, Motte, \& Andre(1999)]{Ward-ThompsonMotteAndre1999}
   Ward-Thompson, D., Motte, F., \& Andre, P. 1999, MNRAS, 305, 143

\bibitem[Willacy \& Williams(1993)]{WillacyWilliams1993}
   Willacy, K., \& Williams, D., 1993, MNRAS, 260, 635

\bibitem[Williams et al.~(1999)]{Williams1999}
  Williams, J., Myers, P., Wilner, D., Di Francesco, J., 1999, Ap J, 513, L61


\bibitem[Zhou et al.~(1994)]{Zhou1994}
  Zhou, S., Evans, N., Wang, Y., Peng, R., Lo, K., 1994, Ap J, 433, 131

\bibitem[Zucconi, Walmsley, \& Galli(2001)]{Zucconi2001}
   Zucconi, A., Walmsley, C., \& Galli, D. 2001, A\&A, 376, 650




\end{thebibliography}
\end{document}